\documentclass[twocolumn,fleqn,,usenatbib]{mnras}
\usepackage{hyperref}
\usepackage{bm}
\usepackage{graphicx}
\usepackage{multirow,tabularx}
\usepackage{enumitem}
\usepackage[flushleft]{threeparttable}
\usepackage[]{color}

\usepackage{natbib}
\usepackage{float}
\usepackage{breqn}
\usepackage[T1]{fontenc}
\usepackage[utf8]{inputenc}

\usepackage{amsfonts,amsmath,amssymb,esint}

\usepackage{color}
\usepackage{xcolor}
\usepackage{url}

\title[X-ray Irradiated AGN Winds]{Time-Dependent AGN Disc Winds I - X-ray Irradiation}

\author[S. Dyda et al.]{
Sergei Dyda,$^{1}$\thanks{sergei@virginia.edu}
Shane W. Davis,$^{1}$
and Daniel Proga$^{2}$
\\
$^{1}$ Department of Astronomy, University of Virginia, 530 McCormick Rd., Charlottesville, VA 22904, USA \\
$^{2}$ Department of Physics \& Astronomy, University of Nevada, Las Vegas, 4505 S. Maryland Pkwy, Las Vegas, NV, 89154-4002, USA
}

\begin{document}

\label{firstpage}
\pagerange{\pageref{firstpage}--\pageref{lastpage}}

\maketitle
\begin{abstract}
We study AGN line driven disc winds using time-dependent radiation hydrodynamics. The key criterion for determining wind launching is the coupling strength of the UV radiation field via the spectral lines of the gas. The strength of these lines in turn relies crucially on the gas ionization state, determined by the local X-ray intensity. 

We consider a suite of models where the central ionizing radiation is affected by scattering, absorption and re-emission by the intervening gas. In a pure attenuation model, the disc launches an episodic wind, as previous studies have shown. Including scattering or re-emission tends to weaken the wind, lowering the mass flux and outflow velocity and if sufficiently dominant, suppressing the outflow entirely. However, the exponential nature of radiative attenuation means only a modest, factor of a few, increase in the absorption cross section can overcome the wind suppression due to scattering and re-emission. We find mass outflow rates of $\sim 20$\% or more of the assumed inflow rate through the disk, indicating that radiation driven winds may significantly alter the structure of the accretion flow. The winds also supply a large, time-varying column of material above the nominal constant disk scale height, which will determine the geometry of reprocessed emission from the central source.  Our results suggest the need for accurate photoionization modeling, radiation transport as well as accretion disc physics, to study their effects on the AGN disc winds.
\end{abstract}

\section{Introduction}
\label{sec:introduction}

Active Galactic Nuclei (AGN) are the central regions of galaxies that exhibit high luminosity and activity, which is attributed to the presence of a supermassive black hole at their centers. One of the most important features of AGNs is the presence of powerful outflows of gas, called AGN winds, that can reach speeds of up to tens of thousands of kilometers per second. These winds can have a range of effects on their host galaxies. They can transport large amounts of energy and momentum away from the central black hole, which can regulate the growth of the black hole and influence the star formation rate in the galaxy. The winds can also heat and ionize the gas in the interstellar medium, which can affect the chemical composition of the gas and the formation of stars. In some cases, the winds can even be powerful enough to expel gas from the galaxy altogether, leading to a decrease in the overall mass of the galaxy. 

 A key unanswered question is what physical processes are responsible for launching and accelerating these strong outflows? Observations of UV absorption lines in AGN spectra has strongly suggested the possibility that winds are accelerated via radiation pressure on lines, so called line driving. For line driving to successfully accelerate the wind two conditions must be met. Firstly the gas in the AGN environment must be sufficiently ionized so that the gas can interact with and gain momentum from the UV radiation field. Secondly, the AGN must provide a high enough UV flux to transfer sufficient momentum to the gas and overcome gravity.

Line driving was originally proposed as a driving mechanism for OB stars ( \cite{Lucy1970}; \cite{CAK1975}, hereafter CAK) and successfully predicted global wind properties such as mass loss rates and outflow velocities. These early successes lead to line driving being applied to study outflows from other compact object system. In the case of cataclysmic variables (CVs), disc wind models (\cite{PSD98}; \cite{PSD99}) have been able to reproduce observed absorption profiles \citep{Proga2003} and the formation of small scale clumps (\cite{DP2018a}; \cite{DP2018b}). In the case of X-ray binaries, the ionization state has been observed to be too high for line driving to be the primary acceleration mechanism, thereby requiring an alternative such as thermal or magnetic driving (\cite{Proga2002}; \cite{Miller2016}; \cite{Higginbottom17}; \cite{Tomaru2020}). The presence of some optical (e.g. \cite{Munoz-Darias2019}; \cite{Mata-Sanchez2018}) and ultraviolet \cite{CastroSegura2022} wind signatures in XRBs
indicates the presence of some lower ionization material, indicating that line driving may play some role in ejecting material. 

Early photoionization studies found the acceleration due to the line force is remarkably constant for stellar temperatures $10 000 \rm{K} \lesssim T \lesssim 50 000$ K \cite{Abbott1982}. The physical environment around AGN is however substantially different than in the OB star context. The gas can be more strongly ionized, and much hotter leading to a severe reduction in the force due to lines. Studying line driven outflows from AGN therefore requires accurate modeling of the ionization state of the gas, the radiation transport and the gas hydrodynamics.

The earliest attempts at photoionization modeling \cite{Stevens1990} parametrized the force multiplier in terms of gas temperature and ionization.  Further efforts were made to alter the orignial CAK formulation due to changes in gas temperature and ionization, and line saturation \cite{Gayley1995}. These force multiplier prescriptions could then be used in analytic (\cite{Arav1994a}; \cite{Arav1994b}) and hydrodynamical modeling (\cite{Murray1995}; \cite{Pereyra1997}; \cite{PSK2000} hereafter PSK00; \cite{PK04} hereafter PK04; \cite{Nomura2016}; \cite{Nomura2017}). More recent attempts  are in the same spirit, \cite{Dannen19}, but rather than parametrizing results, force multiplier data tables are used by hydro simulations directly \citep{Dannen2023}. Further they benefited from substantially more complete data cataloguing line transitions and incident SEDs. The conclusion from these hydrodynamical simulations is that AGN disc winds can be launched and accelerated via line driving, provided the gas is not over-ionized. Either the gas had to be clumpy \cite{Blondin1990}, or some intervening gas had to shield the wind from the central source (\cite{deKool1995}; \cite{Murray1995}).   

In order to further refine these photoionization studies, efforts were made to understand the radiation transfer of the X-rays through the wind. \cite{Schurch2009} used 1D radiation transfer through the PK04 solution to confirm that the viewing angle has a significant effect on spectra due to absorption by the wind. Subsequent development of multi-dimensional Monte-Carlo radiation transfer codes (\cite{Sim2005}; \cite{Sim2008} \cite{Sim2010} hereafter Sim10) allowed for scattering and reprocessing effects to be taken into account. Sim10 showed that the PK04 could reproduce complex spectral features such as broad emission lines, narrow absorption lines and a Compton hump. However, they found that the Monte Carlo derived ionization structure was inconsistent with the simulation's ionization structure. Their work was extended by \cite{Higginbottom14}, who considered longer wavelengths and lower ionization stages
that are critical to allow effective line-driving and the formation of broad absorption lines (BALs). They similarly concluded that once the effects of scattering and reprocessing are included, the ionization state corresponding to PK04 was too high to sustain a line driven wind. Nevertheless, disc wind solutions persist in being able to reproduce observed spectral features when using modern radiation transfer methods (see for example \cite{Matthews2023}).    

Efforts are therefore currently underway to \emph{simultaneously} simulate the radiation transport, gas dynamics and photoionization modeling. Higginbottom et al. (submitted to MNRAS) have used a Monte Carlo radiation-hydro code \citep{Higginbottom2020} to compute line driven disc winds from accreting white dwarfs and found that scattering processes over-ionize the flow and inhibit wind formation. Perliminary work indicates a similar problem will be present in AGN simulations (N. Scepi et al., private communication). Here we use the radiation hydrodynmaics (rad-hydro) code \textsc{Athena++} to study AGN disc winds. We use a two-band model for the radiation, where X-rays evolve according to the time-dependent radiation transport equation whereas the UV field is assumed to be fixed via a geometric optics approximation. The X-rays serve to ionize the gas, which determines the effective number of optically thick lines which couple to the UV field and extract momentum from it. 

In this work, our focus is on the treatment of the X-ray band and its coupling to the gas. We study several coupling regimes between the gas and X-ray field, quantified via the gas opacity. We initially study a pure absorption regime as was first considered in PSK00 and PK04 and recover their non-stationary, disc wind solutions. We then consider scattering effects as studied in Sim10 and show that winds are suppressed due to over-ionization. We further show that re-radiation effects can likewise over-ionize a wind and suppress an outflow. Finally we show that with both scattering and re-radiation operating the wind can still be launched if the absorption opacity are a few times stronger than the electron scattering opacity. 

In the interest of focusing on the question of over/under ionization of the gas due to X-rays we have chosen a correspondingly simplified geometric optics treatment of the UV band. In a future study (Paper II) we will leverage the fully multi-frequency capabilities of \textsc{Athena++} \citep{Jiang2022} to evolve both the X-ray and UV fields using the time-dependent radiation transport equation.

The layout of this paper is as follows. In \S \ref{sec:theory} we describe our simulation set-up, in particular the treatment of the X-ray band which is the novel aspect of this work. In \S \ref{sec:results} we first describe the results of our pure attenuation model which most closely resembles the treatment of previous works. We then describe the effects of including scattering and re-radiation. Finally we conclude in \S \ref{sec:discussion} where we discuss our results, in particular in the context of other treatments of the X-ray radiation using Monte Carlo methods and layout the rational for our future study using time-dependent methods for the UV band.

\section{Theory}
\label{sec:theory}
\subsection{Basic Equations}
\label{sec:equations}
The basic equations for single fluid radiation hydrodynamics are
\begin{subequations}
\begin{equation}
\frac{\partial \rho}{\partial t} + \nabla \cdot \left( \rho \mathbf{v} \right) = 0,
\end{equation}
\begin{equation}
\frac{\partial (\rho \mathbf{v})}{\partial t} + \nabla \cdot \left(\rho \mathbf{vv} + \sf{P} \right) =  \mathbf{G} + \rho \mathbf{g}_{\rm{grav}},
\label{eq:momentum}
\end{equation}
\begin{equation}
\frac{\partial E}{\partial t} + \nabla \cdot \left( (E + P)\mathbf{v} \right) = cG^{0} + \rho \mathbf{v} \cdot \mathbf{g}_{\rm{grav}} + \rho \mathcal{L},
\label{eq:energy}
\end{equation}
\label{eq:hydro}
\end{subequations}
where $\rho$ is the fluid density, $\mathbf{v}$ the velocity, $\sf{P}$ a diagonal tensor with components $P$ the gas pressure. The total gas energy is $E = \frac{1}{2} \rho |\mathbf{v}|^2 + \mathcal{E}$ where $\mathcal{E} =  P/(\gamma -1)$ is the internal energy and $\gamma$ the gas constant. The radiation momentum and energy source terms are $\mathbf{G}$ and $G^0$ respectively and receive contributions from the X-ray (see \S \ref{sec:xray}) and UV bands (see \S \ref{sec:UV}). $\mathcal{L}$ is an optically thin radiative heating term (see \S \ref{sec:heating}) and $\mathbf{g}_{\rm{grav}}$ is the gravitational acceleration (see \S \ref{sec:gravity})

The temperature is $T = (\gamma -1)\mathcal{E}\mu m_{\rm{p}}/\rho k_{\rm{b}}$ where $\mu$ is the mean molecular weight and other symbols have their standard meaning. 

The radiation source terms $\mathbf{G}$ and $cG^0$ are assumed to receive contributions from a UV and an X-ray band 
\begin{subequations}
\begin{equation}
\mathbf{G} = \mathbf{G}_{\rm{UV}} + \mathbf{G}_{\rm{X}}, 
\end{equation}
\begin{equation}
G^{0} = G^{0}_{\rm{UV}} + G^{0}_{\rm{X}}. 
\end{equation}    
\end{subequations}

One can alternatively think of these components as representing emission from a compact central source (the X-ray component) and emission from a disk that is extended in radius (the UV component). In principle, there could be X-rays from the disk and UV from the central source, but the approximation used here is that the ionizing radiation (X-rays) are centrally concentrated and the radiation that potentially drives outflow comes from an extended disk alone. We will therefore model the ionizing central source radiation (X-ray component) via the radiation hydrodynamics module  in Athena++, assuming this component is only incident into the domain on the inner radial boundary.  We will model the disk emission (UV component) as an optically thin radiation force computed using the modified CAK formalism.  

This work is meant to be the first in a series of papers to explore the impact of ionizing radiation on the launching of line driven winds in AGN. In order to connect with earlier work, we have endeavored to match the assumptions and methodological approach of PK04 as nearly as possible. This includes the treatment of the modified CAK line driving prescription, the heating rate, boundary conditions, and gravity, which are described below. As PK04 note, their prescription was a conservative one in that they made a number of unfavorable assumptions (e.g. ignoring the X-ray contribution to the radiation force) that were not conducive to driving an outflow.

The key difference between this work and PK04 is that we leverage the radiation hydrodynamics module of Athena++ to evolve the 
X-ray contribution to the radiation field by solving the time dependent radiation transfer equation directly. In contrast, PK04 solved the radially attenuated X-ray radiation field with an equation of the form
\begin{equation}
    F_x = \exp\left( -\tau \right) \frac{L_x}{4\pi r^2},\label{eq:attenuation}
\end{equation}
where $\tau = \int \rho \kappa dr$ is the optical depth integrated along radial directed rays. We adopt a prescription that allows us to closely emulate this pure attenuation treatment of X-ray flux while also allowing us to include scattered and reradiated radiation, as described below.

\subsection{X-ray Radiation}
\label{sec:xray}

The X-ray radiation field is treated by directly solving the gray (frequency averaged) time dependent radiation transport equation using the implicit implementation in Athena++ \citep{Jiang2021}.  We refer the reader to Jiang (2021) for the specific formulation but the equation solved is equivalent to
\begin{dmath}
\frac{\partial I}{\partial t} + c \mathbf{n} \cdot \nabla I = c S_I, 
\label{eq:dIdt}
\end{dmath}
where $I$ is the frequency integrated specific intensity and the source term
\begin{equation}
    S_I = \Gamma^{-3} \rho \left[ \left( \kappa_P \frac{c a T^4}{4\pi} - \kappa_{E} J_c \right) - \left( \kappa_s + \kappa_{F} \right) \left(I_c - J_c \right) \right],
    \label{eq:sourceterm}
\end{equation}
where $\kappa_s$ is the scattering opacity, $\kappa_{F}$ is the absorption contribution to the flux mean opacity, $\kappa_P$ the Planck mean, $\kappa_E$ the energy mean opacity and $\Gamma = \gamma(1 - \mathbf{n} \cdot \mathbf{v}/c)$ is a frame transformation factor with $\gamma = 1 / \sqrt{1 - v^2/c^2}$ the Lorentz factor. We use a subcript "c" to denote variables evaluated in the comoving frame, including 
\begin{equation}
    J_c = \frac{1}{4\pi}\int I_c \; d\Omega_c,
\end{equation}
the angle averaged comoving frame mean intensity. The X-ray momentum and energy source terms are then
\begin{equation}
    \mathbf{G}_{\rm{X}} = \frac{1}{c}\int \mathbf{n} S_{I} d \Omega,
\end{equation}
\begin{equation}
    cG^0_{\rm{X}} = c \int S_{I} d \Omega.
\end{equation}
The form of equation~(\ref{eq:sourceterm}) is designed so that in the limit $v \rightarrow 0$
\begin{equation}
    \mathbf{G}_{\rm{X}} \rightarrow \frac{\rho  (\kappa_F+\kappa_s)}{c} \mathbf{F}_r,
\end{equation}
\begin{equation}
    cG^0_{\rm{X}} \rightarrow -c\rho (\kappa_P aT^4-\kappa_E E_r).
\end{equation}

For this general scheme outlined above, a common approximation for the opacities is to set $\kappa_E$ equal to $\kappa_P$ so that radiative equilibrium corresponds to $E_r = a T^4$.  Similarly, $\kappa_F$ is often set so that the sum $\kappa_F + \kappa_s$ corresponds to Rosseland mean opacity, with $\kappa_F$ being the contribution from absorption opacity and $\kappa_s$ is set to the electron scattering opacity $\kappa_{\rm es}$, which is nearly constant in the Thomson limit. This scheme is usually a sensible choice {when the radiation variables represent the total frequency integrated radiation field, but here we consider a case where the radiation hydrodynamics solver is only being used to model the X-ray (ionizing) component of the radiation field.  Since X-rays alone do not determine the net heating/cooling or radiation force, we need to consider an alternative prescription that accounts for this complexity.

In our current setup, we choose to turn off the direct coupling of the X-ray radiation field with the gas by setting $\mathbf{G}_{\rm{X}} = c G^0_{\rm{X}} = 0$.  The impact of the radiation on heating/cooling the gas is instead accounted for via the optically thin prescription described in \S ~\ref{sec:heating}.  In principle, we could still retain the momentum coupling given the X-ray contribution to the radiation force, but we drop this term to be consistent with previous work.  Hence, only the radiation force from the UV disc component contributes directly to the radiation force.  Instead, the radiation transfer of the X-rays is included solely to determine the associated ionization parameter
\begin{equation}
    \xi = \frac{4 \pi c E_r}{n},
\end{equation}
where $n = \rho/ \mu m_p$ is the gas number density with $m_p$ the proton mass and $\mu = 0.6$ the mean molecular mass. Since both the heating prescription and the UV line driving force are sensitive to the value of $\xi$, the X-ray component still indirectly impacts the radiation force and radiative heating/cooling.

These considerations also motivate an alternative prescription for the opacities that accounts for the fact that some fraction of the absorbed X-ray radiation field may be thermalized and reemitted outside the X-ray (ionizing) band. Determining the fraction reradiated in the X-rays would require a detailed non-LTE photoionization calculation that is beyond the scope of this work (and current numerical capabilities, at least for fully coupled radiation hydrodynamics).  We intend to explore more sophisticated approximations in future work but for this study we consider a restricted class of models where we parameterize reradiation as a constant fraction $f_\sigma$ of the absorbed X-ray radiation that is remitted in the X-ray band.  We assume this reradiated radiation is isotropically emitted in the comoving frame.

We implement the prescription described above by setting $\kappa_s = \kappa_{\rm es}$, $\kappa_{F} = \kappa_a$, and $\kappa_{E} = \kappa_a( 1 - f_{\sigma})$. Here, $\kappa_a$ is an attenuation opacity that we can vary to see how the radiation field responds under different assumptions. We also set $\kappa_P =0$ to decouple the reemitted radiation from gas temperature, which is instead set by our heating prescription. Our source term then reduces to
\begin{dmath}
S_I = \Gamma^{-3} \rho \left[ \kappa_{\rm es}\left(J_c - I_c \right)  + \kappa_{a} (f_{\sigma} J_c - I_c) \right].
\label{eq:rad_source}
\end{dmath}
With these assumptions $\kappa_a$ is a form of pure absorption opacity that when $f_{\sigma} =0$, simply attenuates the radiation field. When $f_{\sigma} > 0$, a fraction $f_{\sigma}$ of this attenuated radiation is reradiated isotropically in the comoving frame.  When $f_\sigma = 1$ our treatment of absorption and reemission is equivalent to our treatment of scattering.

Finally, we assume a constant radiation flux at the inner radial boundary directed along outgoing radial rays. To mimic the effects of a very compact source at the origin, only the most nearly radial outgoing rays are given a non-zero intensity on the inner boundary.  This is accomplished by using a spherical polar discretization of the unit sphere \citep[see e.g.][]{Whiteetal2023} with a polar axis corresponds to a unit vector in the local $r$ direction and selecting the ring of ray with smallest polar angle. The intensity of the selected rays is chosen such that the total flux in X-rays is a fraction $f_{X} = 0.10 \Gamma_D$ of the disc Eddington fraction.

\subsection{UV Radiation}
\label{sec:UV}
The UV flux is assumed to be time-independent and the wind is assumed to be optically thin to the UV continuum. The momentum transfer can be broken up into contributions from electron scattering and radiation pressure on spectral lines, 
\begin{equation}
\mathbf{G}_{\rm{UV}} = \frac{\rho \kappa_{\rm es}}{c} \int \mathbf{n} I_D f_{UV} \left[ 1 + M(t) \right] d\Omega,
\end{equation}
where the surface integral is over the disc, assumed to be radiating like a Shakura-Sunyaev disc \citep{SS73} with a constant accretion rate $\dot{M}$.  Even though mass is injected into the domain, there is no feedback on the assummed mass flow rate, which remains constant.  Hence, the radiation at the disk surface is modeled as
\begin{dmath}
I_D = \frac{3}{\pi}\frac{GM_{BH}}{r_{*}^2}\frac{c}{\kappa_{\rm es}} \Gamma_D \left[ \left(\frac{r_{*}}{r}\right)^3 \left( 1 - \left[\frac{r_{*}}{r}\right]^{1/2} \right) + \frac{x}{3\pi} \left\{ \sin^{-1}\left( \frac{r_{*}}{r} \right) - \frac{r_{*}}{r} \left( 1 - \left[ \frac{r_{*}}{r}\right]^2 \right)^{1/2} \right\} \right].
\label{eq:SSIntensity}
\end{dmath}
where $r_*=6 GM/c^2$, $x$ is the re-radiation factor ($x=1$ in this work), and $\Gamma_D = \dot{M}/\dot{M}_{\rm Edd}$, where
\begin{equation}
\dot{M}_{\rm Edd} = \frac{4 \pi G M_{BH}}{\eta \kappa_{\rm es} c}.
\end{equation}
We assume an efficiency $\eta=1/12$ in this work. In this model, the spectrum is blackbody, with a temperature
\begin{equation}
    T_D = \left[ \frac{\pi I_D}{\sigma} \right]^{1/4}.
    \label{eq:SSTemp}
\end{equation}
The fraction of the intensity in the UV is then found by integrating the Planck function 
\begin{equation}
    f_{UV} = \frac{\pi}{\sigma T_D^4} \int_{\lambda_1}^{\lambda_2} \frac{2hc^2}{\lambda^5} \frac{1}{ \exp \left\{hc / \lambda k_b T_D \right\}- 1},
\end{equation}
where $\lambda_1 = 200$ \AA \ and $\lambda_2 = 3 200$ \AA \ and $\sigma$ is the Stefan-Boltzmann constant.

The strength of the spectral lines, relative to electron scattering is quantified via the force multiplier \cite{Owocki1988}
\begin{equation}
M(t) = kt^{\alpha} \left[ \frac{\left(1 + \tau_{\rm{max}}\right)^{1 - \alpha} - 1}{\tau_{\rm{max}}^{1 -\alpha}}\right],
\end{equation}
with $\alpha = 0.6$ the ratio of optically thick to optically thin lines, 
\begin{equation}
k = 0.03 + 0.385 \ \exp \left\{ - 1.4 \ \xi^{0.6} \right\},
\end{equation}
the number of lines and $\tau_{\rm{max}} = t \eta_{\rm{max}} $ with the parameter 
\begin{equation}
\log_{10} \eta_{\rm{max}} = 
 \begin{cases} 
       & 6.9 \exp \left\{ 0.16 \ \xi^{0.4} \right\} \\
      & 9.1 \exp \left\{ - 7.96 \times 10^{-3} \ \xi \right\} \\ 
   \end{cases}
\end{equation}
determining the maximum number of lines available and the optical depth parameter 
\begin{equation}
t = \frac{\kappa_{\rm es} \rho v_{\rm{th}}}{\left| dv_{l} / dl\right|},
\end{equation}
with the thermal velocity of the gas $v_{\rm{th}} = 4.2 \times 10^5$ cm/s. The disc as a source of radiation is taken to extend outside the computational domain to
$R_d = 1500 r_g$.

Finally, the work done by the UV radiation force is then
\begin{equation}
cG^{0}_{\rm{UV}} = v \cdot \mathbf{G}_{\rm{UV}}. 
\end{equation}

\subsection{Radiative Heating}
\label{sec:heating}
We assume a locally optically thin radiative heating and cooling due to Compton, X-ray photoionization, Bremsstrahlung and lines. For a 10keV Bremsstrahlung, \cite{Blondin1994} found a heating function of the form
\begin{equation}
    \rho \Lambda = n^2 \left( G_{\rm{Comp}} + G_{\rm{Xray}} - L_{Brem} - L_{\rm{lines}}\right),
\end{equation}
 where the contributions in cgs units, $\left[ \rm{erg \ cm^{-3} \ s^{-1}} \right]$,  due to Compton, X-ray photoionization, Bremsstrahlung and lines are respectively
 \begin{subequations}
\begin{equation}
G_{\rm{Comp}} = 8.9 \times 10^{-36} \xi \left( T_X - T \right), 
\end{equation}
\begin{equation}
G_{\rm{Xray}} = 1.5 \times 10^{-21} \xi^{1/4} T^{-1/2} \left( 1 - T/T_X \right),
\end{equation}
\begin{equation}
L_{\rm{B}} = 3.3 \times 10^{-27} T^{1/2}, 
\end{equation}
\begin{equation}
\begin{split}
L_{\rm{lines}} = \delta \Big[ 1.7 \times 10^{-18} \exp \left\{ -\frac{T_{L}}{T}\right\} \xi^{-1} T^{-1/2}  + 10^{-24} \Big],
\end{split}
\end{equation}
 \end{subequations}
with the line temperature $T_L = 1.3 \times 10^{5}$ and $\delta = 1$ the parameter that controls the efficiency of line cooling.

\subsection{Gravitational Potential}
\label{sec:gravity}
We treat the disc as a mass reservoir of fixed density and fixed radiation intensity. Thus, we are implicitly assuming that our disc boundary is at a height $h$ above the disc midplane. For radiation dominated discs \citep{SS73} the photosphere should be roughly a few times the Eddington fraction in units of $r_{\rm{ISCO}}$. We take the shift to be 
\begin{equation}
h = 3 \Gamma_D r_* f_h,
\end{equation}
above the disc midplane. In this work we choose $f_h =0.85$, where $f_h=1$ would correspond to a displacement by one Shakura \& Sunyaev scale height. In these shifted coordinates, the gravitational acceleration acquires additional non-radial components
\begin{subequations}
    \begin{equation}
        g_r = \frac{GM_{\rm{BH}}}{r^{'2}} \sin \alpha,
    \end{equation}
        \begin{equation}
        g_{\theta} = \frac{GM_{\rm{BH}}}{r^{'2}} \cos \alpha,
    \end{equation}
\end{subequations}
where
\begin{subequations}
    \begin{equation}
        \sin \alpha = \frac{h \sin \theta}{r'},
    \end{equation}
        \begin{equation}
        \cos \alpha = \frac{r + h \cos \theta}{r'},
    \end{equation}
\end{subequations}
and
\begin{equation}
    r^{'2} = r^2 + 2 r h \cos \theta + h^2.
\end{equation}

\subsection{Simulation Parameters}
Follwoing PK04 we consider a central black hole mass $M_{BH} = 10^{8} M_{\odot}$, corresponding to an innermost stable circular orbit (ISCO) $r_* = 6 r_g = 8.8\times 10^{13}\rm{cm}$. We report some results in terms of gravitational radii $r_g = GM/c^2 = 1.4 \times 10^{13} \rm{cm}$ and the orbital period at the inner boundary $t_0 = 2\pi ((60 r_g)^3/GM)^{1/2} = 1.44 \times 10^{6} s$

We impose inflow (outflow) boundary conditions at the inner (outer) radial boundaries and axis boundary conditions along the $\theta$ = 0 axis. We assume a reflection symmetry about the $\theta = \pi/2$ midplane.
We use a vacuum boundary condition for the radiation along the disc midplane and outer radial boundaries and keep the X-ray flux fixed at the inner radial boundary. After every full time step we reset $\rho_d = 10^{-8} \rm{g cm^{-3}}$, $v_r = 0$ and $v_{\phi} = v_K = \sqrt{GM/r}$. We also impose that the vertical velocity component $v_{\theta}$ is unchanged due
to resetting density.

We choose a domain size $n_r$ × $n_{\theta}$ = 96 × 140. The radial domain extends over the range $10 \ r_{*} < r < 500 \ r_{*}$ with logarithmic spacing $dr_{i+1}/dr_{i} = 1.05$. The polar angle range is $0 < \theta < \pi/2$ and has logarithmic spacing $d \theta_{j+1}/d \theta{j} = 0.938$ which ensures that we have
sufficient resolution near the disc midplane to resolve the acceleration of the flow.

Initially, the cells along the disc are set to have $\rho = \rho_d$, $v_r = v_{\theta} = 0$, $v_{\phi} = v_K$. In the rest of the domain $\rho = 10^{-20} \ \rm{g cm^{-3}}$ and $v_r = v_{\theta} = v_{\phi} = 0$. Everywhere the temperature is constant along vertical cylinders corresponding to the Shakura-Sunyaev disc temperature at the base given by \cite{SS73}.

We take the disc Eddington number $\Gamma_D = 0.5$ and re-radiation factor $x = 1$ (see eq \ref{eq:SSIntensity}). This choice of $\Gamma_D$ corresponds to an accretion rate $\dot{M} = 1.3 \; \rm M_\odot/yr$ for $M_{BH}=10^8 M_\odot$. We assume the luminosity of the central source radiation in X-rays is 10\% of the corresponding disk luminosity.  We assume the central source does not emit in the UV. The UV radiation from the disc is assumed to extend all the way from the ISCO (effectively outside the simulation domain) with $r_{*} \leq R_d \leq 1500 r_{*}$. A sphere of radius $r_{*}$, effectively the black hole and nearby corona, is assumed to be optically thick for purposes of shielding the wind from the backside of the disc.  
. 
We impose a density floor $\rho_{\rm{floor}} = 10^{-22} \rm{g cm^{-3}}$ which adds matter to stay above this floor, while conserving momentum, if the density ever drops below it. In addition, we have a temperature floor at the same cylindrical radius.

\section{Results}
\label{sec:results}

\begin{table*}
\centering
\begin{tabular}{ l l c c c c c c c c l}
\hline \hline
\multirow{2}{*}{Model} & \multirow{2}{*}{} & \multicolumn{3}{c}{Opacity [$\kappa_{\rm{es}}$]} & \multicolumn{4}{c}{Wind Properties} & Description \\
& & $\kappa_a$  & $\kappa_s$ & $f_{\sigma}$  & $\log \dot{m} \ [M_{\odot}/\rm{yr}]$ & $v_{\rm{out}} $ [km/s] & $\omega$ & $\Delta t_{\rm{high}} \ [t_0]$ \\ \hline \hline
A1 & Pure attenuation & 1 & 0 & 0  & -0.5 & 11 000 & $76^{\circ}$ & -- & Episodic Wind \\
S1 & Pure scattering & 0 & 1 & 0  & -- & -- & -- & -- & No wind\\ \hline
A1S1 & Attenuation \& Scattering & 1 & 1 & 0   & -1.3 & 6 000& $69^{\circ}$ & 85 & Episodic Wind\\
A2S1 & Attenuation \& Scattering & 2 & 1 & 0   & 0.53 & 10 000 & $60^{\circ}$ & -- & Stationary wind\\
A10S1 & Attenuation \& Scattering & 10 & 1 & 0   & 1.1 & 20 000 & $70^{\circ}$ & -- & Stationary wind \\ \hline
A1S1f50 & Scattering \& 50\% re-emission & 1 & 1 & 0.5   & -- & -- & -- & -- & No Wind \\
A2S1f50 & Scattering \& 50\% re-emission  & 2 & 1 & 0.5   & -1.4 & 6 000 & 64 & 102 & Episodic wind \\ \hline
A1f5 & 5\% re-emission & 1 & 0 & 0.05  & -1.7 & 9 000 & $70^{\circ}$& 92 & Episodic wind \\ 
A1f10 & 10\% re-emission & 1 & 0 & 0.10  & -1.6  & 5000 & $67^{\circ}$ & 107 & Episodic wind\\
\hline\hline
    \end{tabular}
\caption{Summary of models listing the gas/radiation coupling parameters that were varied (absorption opacity $\kappa_a$, scattering opacity $\kappa_s$ and re-emission fraction $f_{\sigma}$) and the resulting wind properties including mass outflow rate $\dot{m}$, outflow velocity $v_{\rm{out}}$, inclination angle $\omega$ and time between high state outflows $\Delta t_{\rm{high}}$.}
\label{tab:summary}
\end{table*}

\subsection{Absorption Effects}

\begin{figure*}
    \centering
    \includegraphics[scale=0.8]{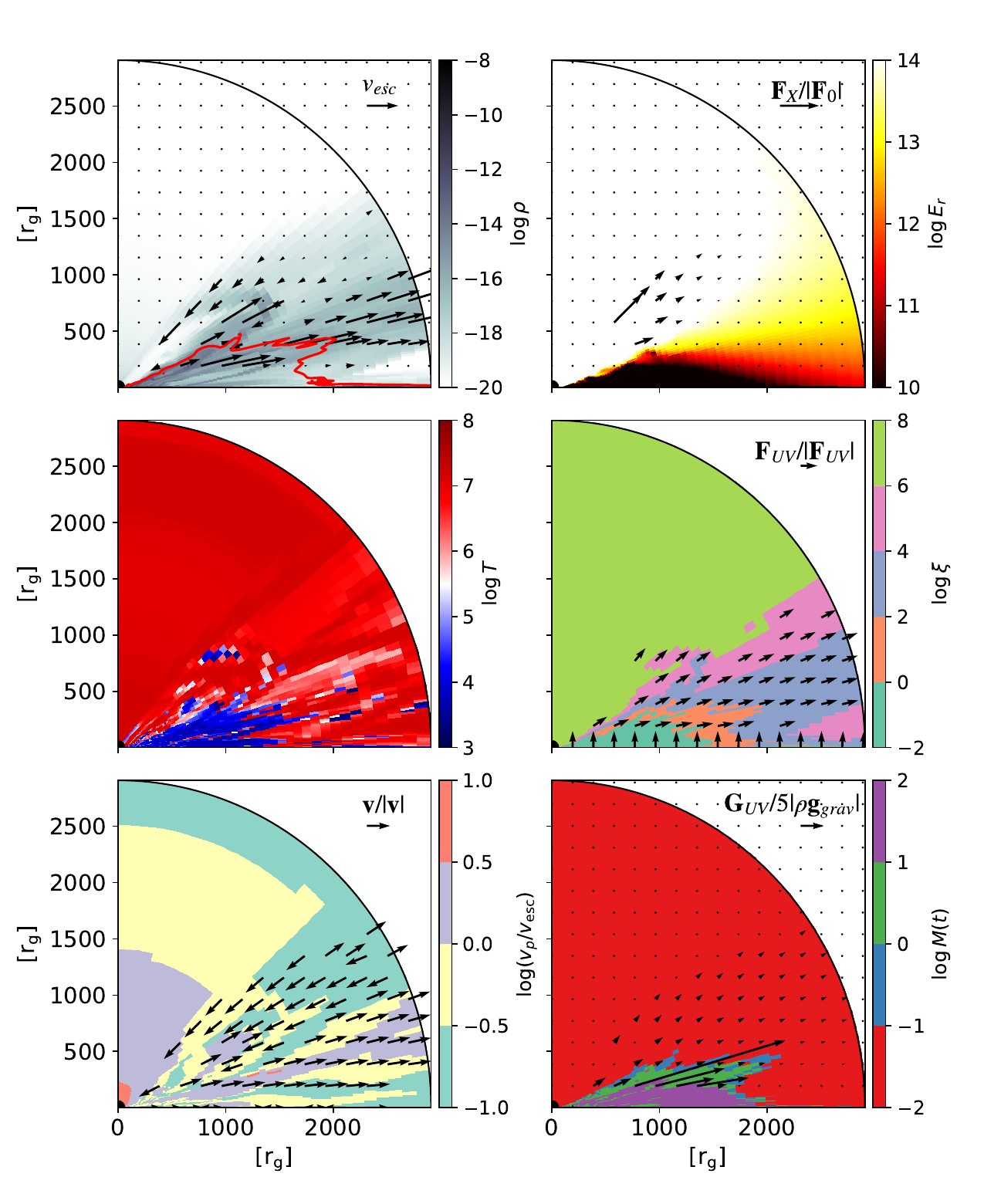}
    \caption{Wind solution at $t = 196 t_0$ showing density (and gas velocity vectors and $M(t) = 2$ contour in red), temperature, poloidal velocity (and unit gas velocity), force multiplier (and radiation force vectors), ionization parameter (and unit UV flux vectors) and X-ray radiation energy density ( and X-ray flux). We only plot vectors in parts of the domain above a representative density floor of $\rho = 10^{-19} \ \rm{gcm^{-3}}$ for clarity. We see a strong outflow at $\omega \sim 75^{\circ}$, with acceleration primarilly occuring at the base of the wind where the ionization is low, $\xi \lesssim 2$ and the force multiplier correspondingly large, $M(t) \sim 10$}
    \label{fig:4panel}
\end{figure*}

\begin{figure*}
    \centering
    \includegraphics[scale=0.45]{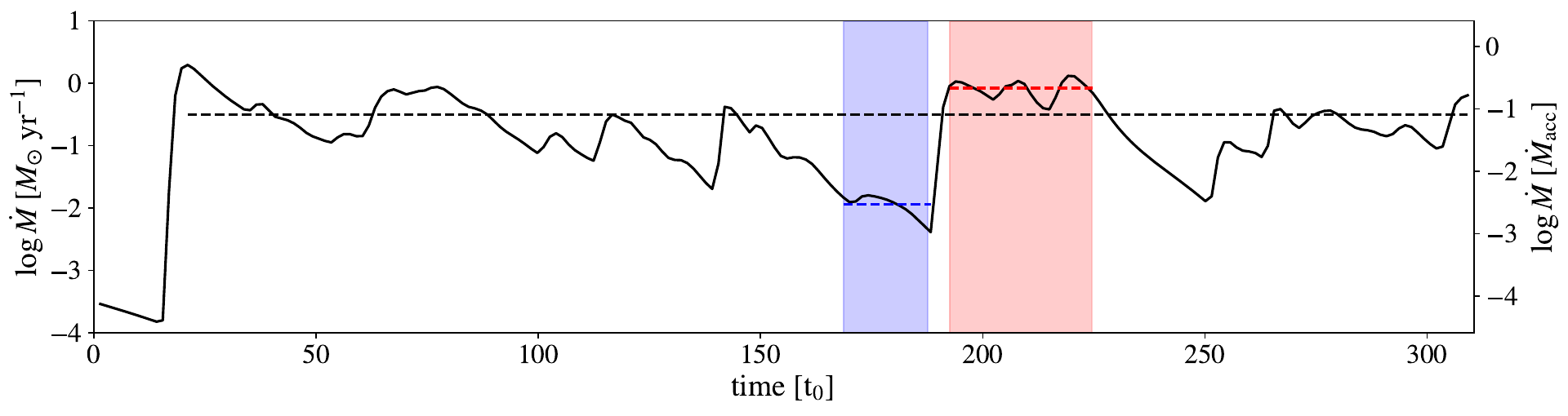}
\includegraphics[scale=0.45]{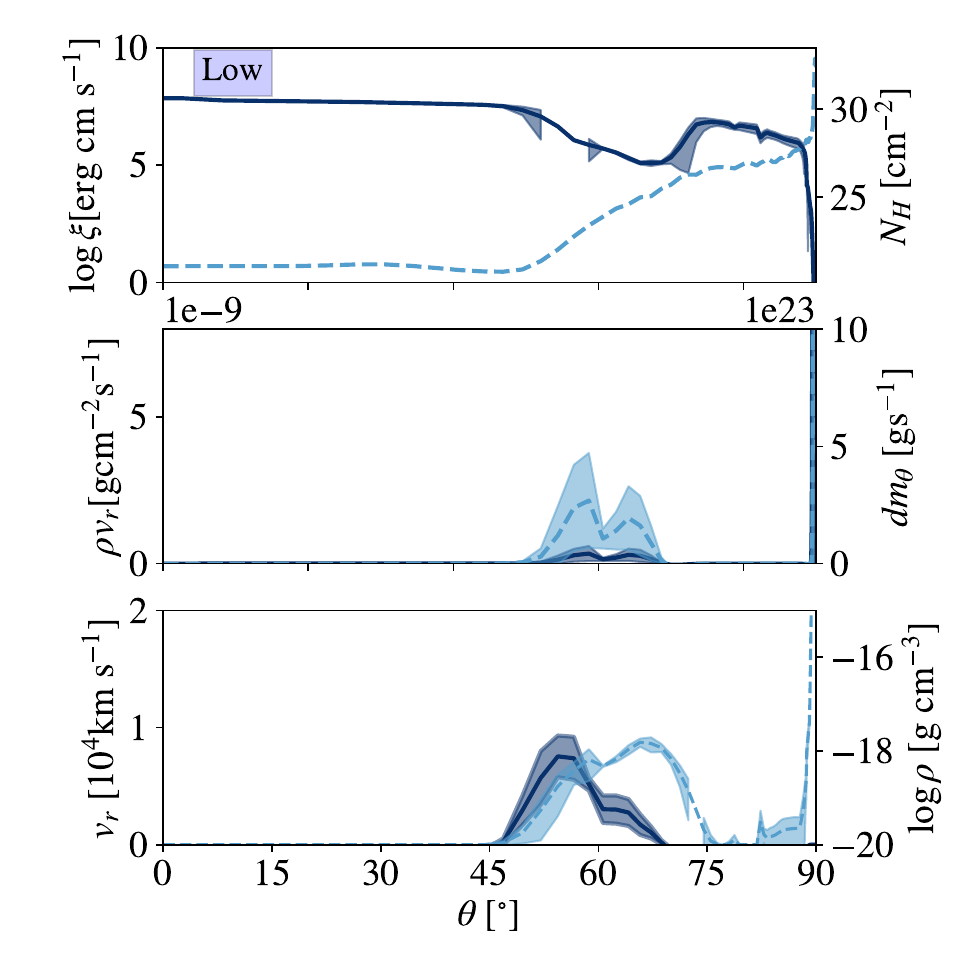}
    \includegraphics[scale=0.45]{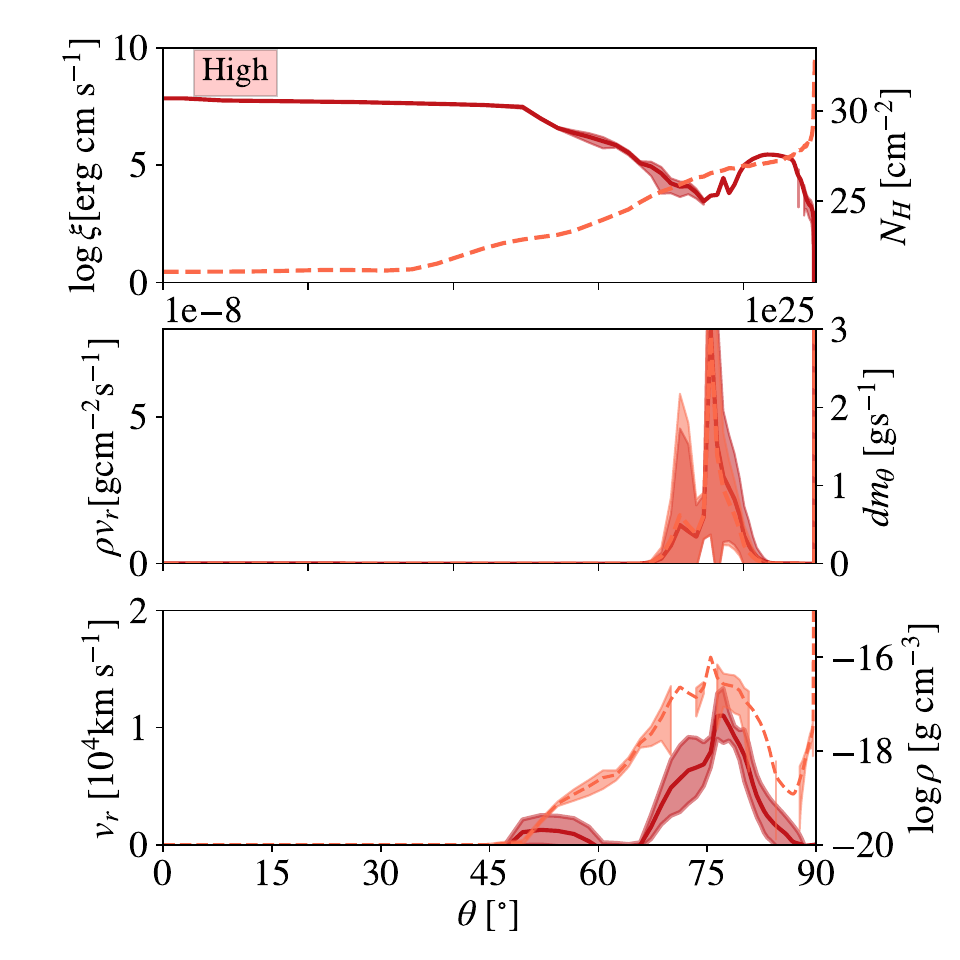}
    \includegraphics[scale=0.95]{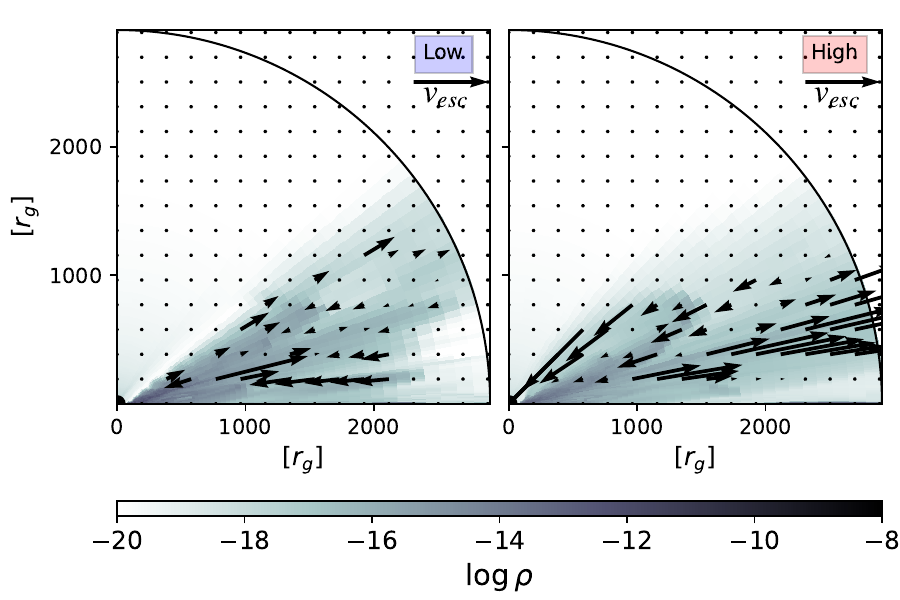}
    \caption{Model A1. \textit{Top panel} - Time dependent mass flux through the outer boundary for $0^{\circ} \leq \theta \leq 85^{\circ}$. We highlight a typical low (high) state in blue (red) where the wind is weak (strong). \textit{Center panels} - Dynamical wind variables for the low (left) and high (right) states. The shading shows the one standard deviation variations of the respective quantities with respect to time. The low state is dominated by a fast, low density component with moderate inclinations. The high state has a slower, less dense component at high inclinations and a fast, dense component at low inclinations that is responsible for the enhanced mass outflow rate. \textit{Bottom panels} - Time-averaged density and velocity fields for the low (left) and high (right) states. The high state shows evidence of a failed wind at moderate inclination angles.}
    \label{fig:A1out}
\end{figure*}
For our fiducial model, we utilize a pure attenuation ($\kappa_a=\kappa_{\rm es}$ and $f_\sigma =0$) treatment to emulate equation~(\ref{eq:attenuation}). We first test our results by using an angular grid with a single radial ray. This single ray approximation is the closest we come to equation~(\ref{eq:attenuation}) with the primary difference being that $\tau$ is accumulated over the light crossing timescale to the current radius rather than constructed from the current instantaneous value of the column. The wind properties obtained from this single ray approximation were consistent with PK04 results.

Since we are ultimately interested in the effects of scattering and reemission, we perform our simulations with $n=256$ rays uniformly covering the unit sphere. We checked for convergence of our model using $n_r = 128, 512$ rays and concluded this was a sufficiently resolved angular resolution.

We find radial, time variable disc wind solutions, broadly in agreement with the results in PK04. In Fig. \ref{fig:4panel} we show (going counter-clockwise)  density (and gas velocity vectors), temperature, poloidal velocity (and unit gas velocity), force multiplier (and radiation force vectors), ionization parameter (and unit UV flux vectors) and X-ray radiation energy density ( and X-ray flux) at $t = 196 t_0$. On long time-scales, the outflow has two components, a faster, less dense flow at $\theta \sim 50^{\circ}$ and a denser, slower component at $\theta \sim 75^{\circ}$. 

The fast component has a density $\rho \sim 10^{-18} \rm{g \ cm^{-3}}$ and velocities as high as 11 000 km/s. The slow component is an order of magnitude denser with $\rho \sim 10^{-17} \rm{g \ cm^{-3}}$ and velocities of $v \sim 5 \ 000$ km/s.

The wind is optically thick with $N_H \gtrsim 10^{24}$ for $\theta \lesssim 60^{\circ}$. We can understand the wind launching via the X-ray radiation energy density. In the wind, $E_r$ is low, ensuring the ionization is also sufficiently low and the force multiplier large enough for the line driving to overcome gravity. From the force multiplier plot, we see the radiation force is sufficiently strong ($M(t) > 1$, indicated by the green contours) in the very base of the disc, close to the central region. Within this contour the ionization $\xi \lesssim 10^4$ is sufficiently low so as to not suppress the force multiplier. In the limit of an optically thin wind, $\tau_{\rm{max}} \rightarrow 0$, the maximum force multiplier $M_{\rm{max}} = 2$ for $\log \xi = 2.23$. For our Eddington parameter $\Gamma = 0.5$, this is roughly the minimum force multiplier to launch a wind, so our best case scenario is wind launching in the orange colored contours of the ionization plot.

The flow is time variable and we see periods with stronger and weaker outflows. In the top panel of Fig \ref{fig:A1out} we show the total mass flux out the outer boundary in $0^{\circ} \leq \theta \leq 85^{\circ}$. We excised the gas close to the disc because the disc density is quite high relative to the wind and dominated the signal. We have identified two epochs representative phases of the wind, a ``weak" state (shaded in blue) and a ``high" state (shaded in red). For each epoch, we show the time-averaged wind fluxes exiting the outer boundary as a function of $\theta$ (middle panels). We plot the column density $N_H = \int n dr$, the momentum flux $\rho v_r$ and the mass flux density $dm_{\theta} = 4 \pi \rho v_r r^2 d \theta$.  The shading shows the one standard deviation variations of the respective quantities with respect to time. The lower panels show the time-averaged density and velocity field.  

During the low state, mass leaves the domain only at $\theta \sim 60^{\circ}$. During the high state, the mass flux is dominated by the more radial component where the velocities reach $\gtrsim 11 \ 000 \ \rm{km/s}$ but there is also a slower, less dense component at higher latitudes. The density profiles show that even in the high state the flow is unsteady. We see evidence of a failed wind with matter accreting both at latitudes just above the slow part of the wind.

Close to the disc, where the gravitational force scales as $|\mathbf{g}_z| \approx GMz/r^3$ matter can be supported against gravity via the radiation force due to electron scattering, which is approximately constant near the photosphere. As the matter is launched, the relative contribution of gravity and electron scattering changes and further acceleration requires line driving to take over. If line driving is too weak, due to overionization for example, the wind will fail and the launched gas will fall back towards the central object. The wind launching is a highly dynamic process, with gas being launched, failing, and then being swept back up in the outflow.

In Table \ref{tab:summary} we show a summary of our other wind models. We indicate the gas/radiation coupling parameters that were varied (absorption opacity $\kappa_a$, scattering opacity $\kappa_s$ and re-emission fraction $f_{\sigma}$) and the resulting wind properties including mass outflow rate $\dot{m}$, outflow velocity $v_{\rm{out}}$, inclination angle $\omega$ and time between high state outflows $\Delta t_{\rm{high}}$. The goal is to explore the effects of scattering and re-radiation on the fiducial attenuation model. As an initial condition, we use the wind from the fiducial run at $t = 2 \times 10^{8} \rm{s} \approx 140.5 t_0$, where a wind has been well established. We list the main radiation variables varied, namely the absorption and scattering opacitites and the re-radiation fraction. For each set of parameters, the solution is described as either a stationary wind, an episodic wind or having no wind. In the case of an outflow, we list the time averaged mass outflow rate, the peak outflow velocity and the azimuthal position of this velocity peak. In the case of episodic winds, the wind parameters are expressed during the ``high" state during which an outflow is occurring. For episodic winds we also list the average period between outflow maxima.  

\subsection{Scattering Effects}

\begin{figure*}
    \centering
    \includegraphics[scale=0.45]{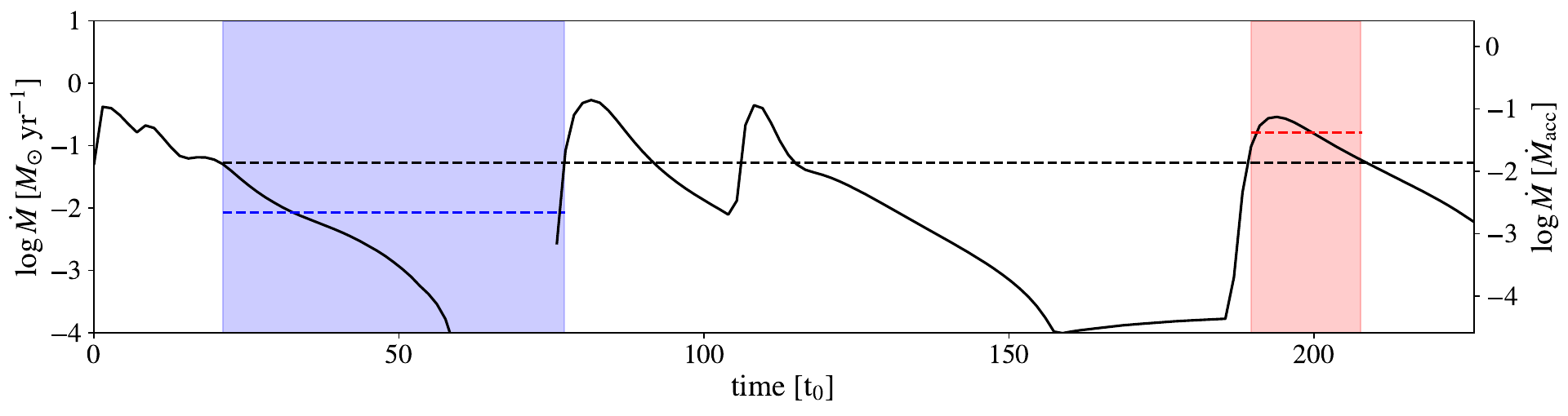}
    \includegraphics[scale=0.45]{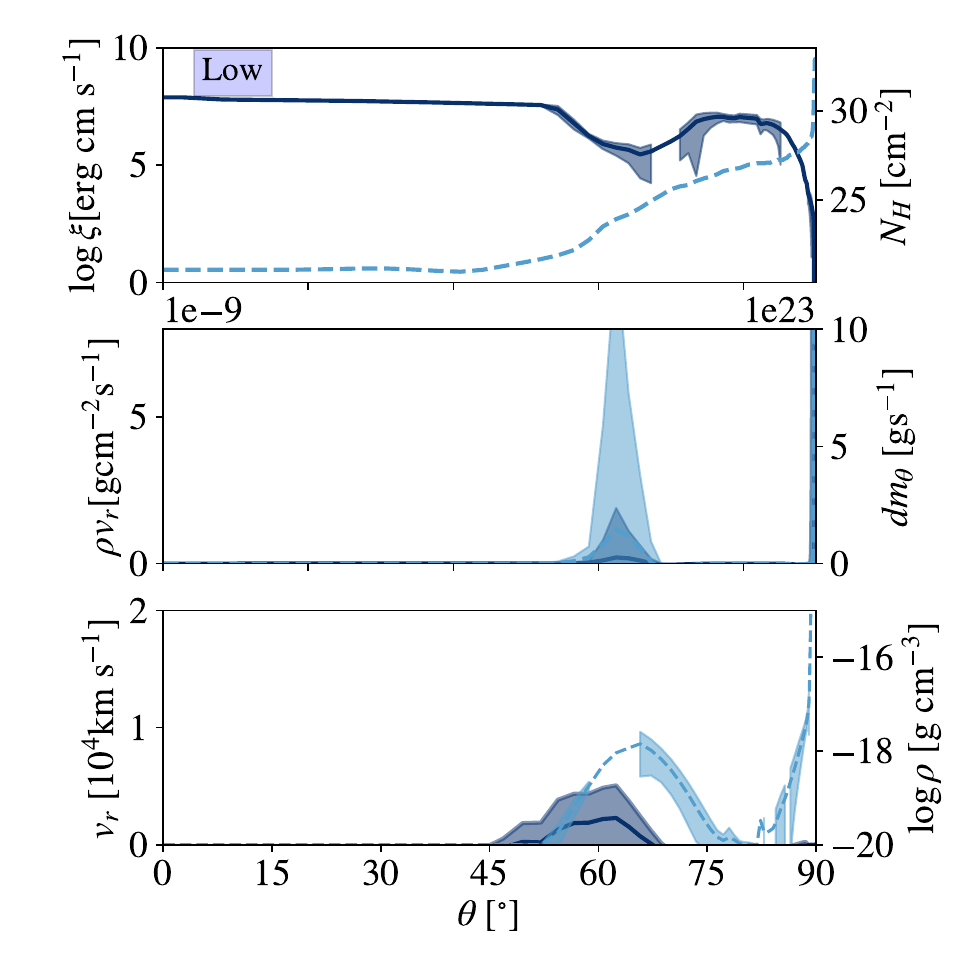}
    \includegraphics[scale=0.45]{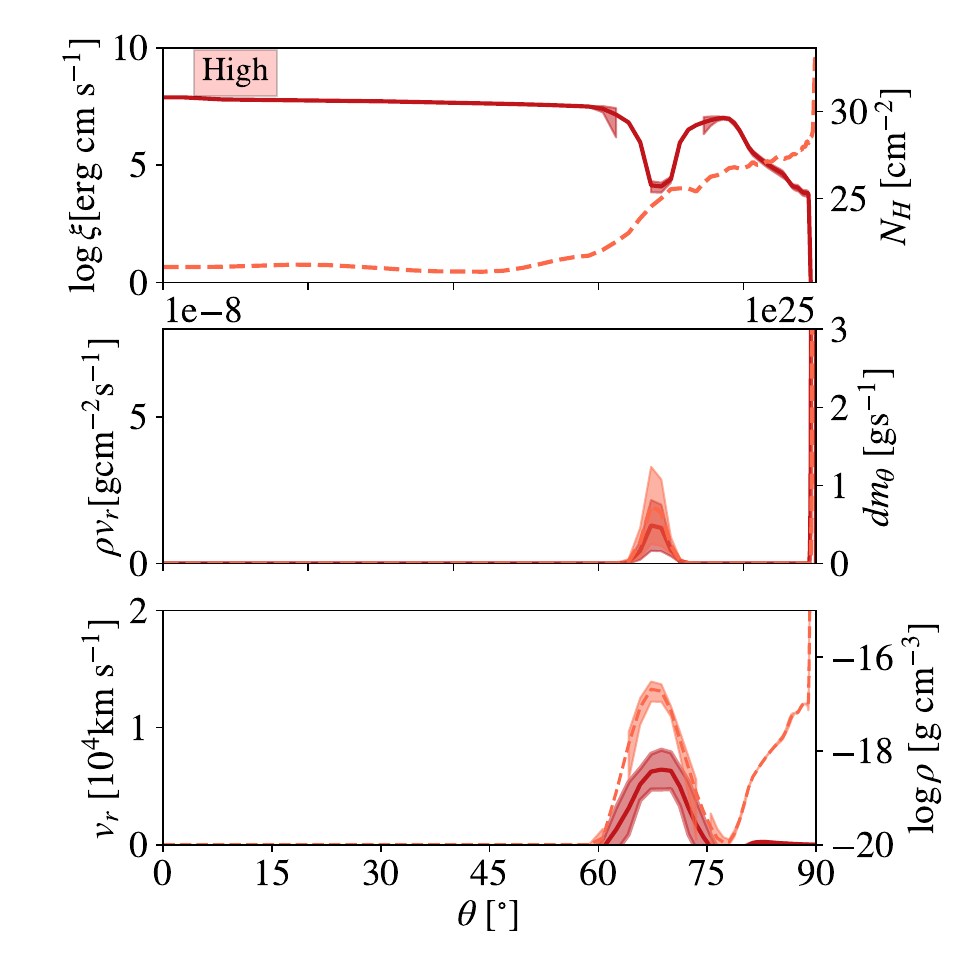}
    \includegraphics[scale=0.95]{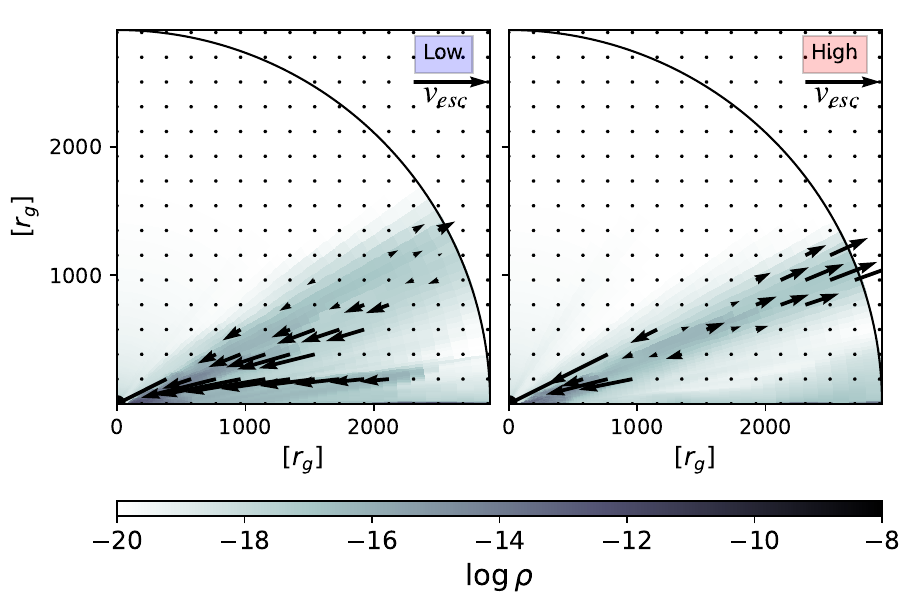}
   \caption{Same as Fig \ref{fig:A1out} but for model A1S1.}
    \label{fig:A1S1out}
\end{figure*}

We next consider cases in the pure scattering limit, where $\kappa_a = \kappa_P = 0$, $f_{\sigma} = 0$ and $\kappa_s = \kappa_{es}$. The radiation field changes on timescales of $\sim 10 \ t_0$ as radiation is scattered from the low density funnel into the higher density wind. In the wind launching region, the X-ray energy density increases by $\sim 4$ orders of magnitude, leading to a commensurate increase in the ionization parameter. This effectively turns off the line driving and the wind gradually fails as new gas does not launch and the already launched gas exits the simulation domain.

To facilitate wind launching, we try the alternate scenario where matter is allowed to enter the initially empty domain from the disc midplane, as we used to launch our fiducial run. Matter enters the simulation domain, forming a thin disc, but the X-ray energy density is even higher than when using our restart prescription. Our results suggest that launching AGN disc winds is not possible in a pure scattering regime. 

We therefore turn our attention to a regime with both scattering and absorption, $\kappa_a = \kappa_s = \kappa_{es}$. Unlike our fiducial model, we find a quasi episodic wind. In the top panel of Fig \ref{fig:A1S1out} we plot the mass flux exiting the simulation domain as a function of time. We identify several outbursts and periods of quiescence in the outflow, and highlight one such low (high) state shaded in blue (red). For these respective states we plot the time-averaged values of the dynamical variables at the outer boundary, and the density and velocity fields. As with the fiducial model, the low state is characterized by a slower, less dense and higher inclination flow than the high state. 

The quasi periodicity is due to the ionization structure of the launching region. After outburst, the wind is too ionized and gas accretes towards the central region. This causes the density near the inner radius to peak, and the ionization to decrease. This allows for a larger force multiplier and a new outflow is launched. The time scale of this process is thus set by a combination of the inflow time and outflow time of the accreting/outflowing gas.  

Increasing the absorption opacity has the effect of making the wind stationary and strengthening the outflow. Increasing $\kappa_a = 2 \kappa_{es}$ causes the wind to become stationary, with a fourfold increase in the mass flux compared to the quasi episodic winds high state and a doubling of the peak velocity. Further increasing $\kappa_a = 10 \kappa_{es}$, the values explored in PSK00, yield winds with 15 times the mass flux and peak velocities of 20 000 km/s.

\subsection{Re-radiation Effects}
We consider models where a fraction of absorbed radiation is re-emitted in the X-rays with $\kappa_{E} = \kappa_a \left( 1 - f_{\sigma} \right)$ and $0 < f_{\sigma} \leq 1$.

Firstly consider models with only an absorption opacity. With only 5\% re-radiation, $f_{\sigma} = 0.05$, we find a very weak, episodic wind. Qualitatively it resembles the $A1S1$ model, with slightly weaker $\sim 50\%$ lower mass flux but 50\% greater peak outflow velocities. Increasing the re-radiation to 10\%, the qualitative behaviour is unchanged and we again find a weak, episodic outflow with peak velocities of $\sim 5 \ 000 \rm{km/s}$. Unsurprisingly, allowing gas to re-radiate has a similar physical effect to the wind as scattering. The X-rays tend to isotropize, and penetrate into the wind acceleration region. This tends to weaken the wind as gas becomes over-ionized. Further increasing the re-radiation fraction eventually leads to the wind being fully suppressed, which we see for $f_{\sigma} = 0.5$.

We now consider models with both scattering, and absorption opacity. We set the re-radiation to 50\%. When $\kappa_a = \kappa_s = \kappa_{es}$ we find that no wind is launched. This is expected as the model with these parameters and no scattering failed to launch a wind, so scattering could serve only to further inhibit wind acceleration. 

We therefore increase the absorption opacity to $\kappa_a = 2 \kappa_{es}$ and find that we again enter a regime of episodic outflows. The wind is both quantitatively and qualitatively similar to the model without re-emission, A2S1.

\section{Discussion}
\label{sec:discussion}

\begin{figure*}
    \centering
    \includegraphics[scale=1.0]{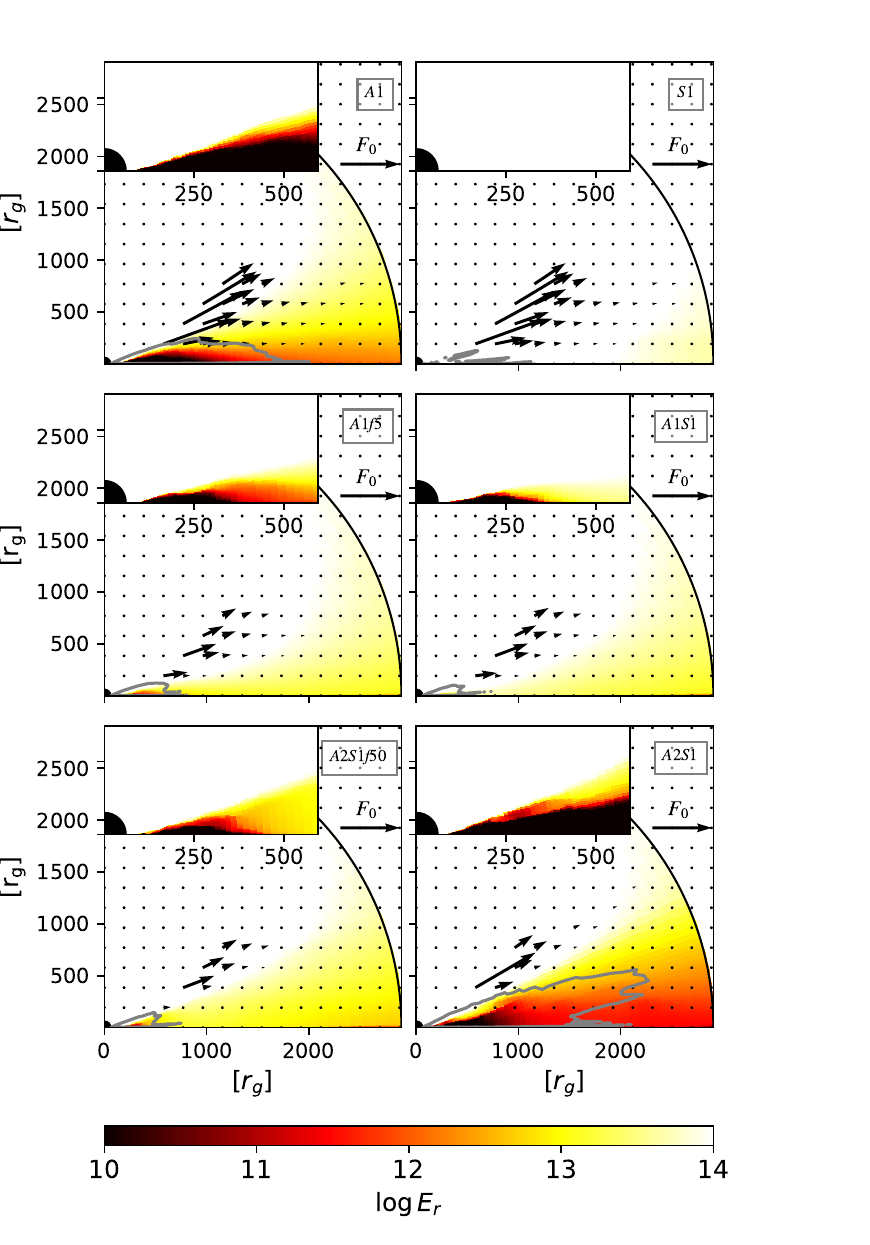}
    \caption{Time averaged radiation field for different disc wind models. The grey contour indicates where $M(t) = 2$, where we expect the radiation force due to lines to overcome gravity.The radiation flux is only plotted for regions with $\rho > 10^{-17} \ \rm{gcm^{-3}}$ Models with an extended region of low X-ray radiation density (A1, A2S1) are able to sustain stronger. Winds with only a small such region (A1f5, A1S1, A2S1f50) can only launch episodic winds when failed winds shield the innermost part of the disc. Model S1 is always over-ionized and cannot launch winds.}
    \label{fig:rad_field_summary}
\end{figure*}

We have used time-dependent radiation transfer to reproduce well established results for AGN disc winds that accounted only for radiative attenuation. We then use our more sophisticated radiation treatment to study the effects of scattering and re-radiation. We have used an idealized treatment of radiative absorption, scattering and reprocessing to develop an intuition for their relative effects on radiation driven disc winds. 

Our study demonstrates that radiative absorption of X-ray photons must be sufficiently strong to prevent over-ionization and allow wind launching. If scattering and re-radiation effects are neglected, absorption opacities equal to electron scattering is sufficient. In a pure scattering regime no wind is launched. If scattering and absorption opacities are equal, we find a much weaker, episodic outflow. The time scale of the variability is set by the dynamical time for the wind to fail, fall back towards the black hole and shield the outer parts of the disc to allow for a new outflow to be launched. If we allow re-radiation within the X-ray band, a modest 5\% re-radiation fraction makes the wind become episodic as in the scattering case. If re-radiation fraction is 50\% we require an enhancement of a few times electron scattering to the absorption opacity to maintain a wind.

These results show that absorption and re-processing of X-rays is crucial to launch an AGN line driven disc wind. Any process which enhances the ionization parameter by increasing the X-ray energy density such as scattering or re-emission will tend to weaken and ultimately suppress the wind. In Fig. \ref{fig:rad_field_summary} we plot the time-averaged X-ray energy density and radiation flux in units of the central object $F_0 = L_{BH}/4\pi r_0^2$. The grey contour indicates the $M(t) = 2$ surface, interior to which we expect the force due to line driving $f_{LD} \sim \Gamma M(t) > f_{\rm{grav}}$. We see that the models with stationary winds, A1 and A2S1 have a geometrically large region where the radiation force dominates, whereas models with periodic outflows (A1f5, A1S1, A2S1f50) have only a geometrically small region. The model with no wind, S1 has no such region. This region of enhanced launching has low radiation energy density, consistent with a low ionization parameter.  

Fortunately, attenuation effects scale like $\exp \left( - \rho \kappa dr \right)$, so a modest, factor of a few, increase in the opacity can exponentially suppress X-ray photons. Thus, while we find that for $\kappa_a = \kappa_{es}$ a mere 5\% re-radiation effect causes the wind to become weak and episodic and 50\% re-radiation suppresses the wind, simply increasing the absorption opacity to $\kappa_a = 2 \kappa_{es}$ allows for an episodic wind with 50\% re-radiation. Photoionization studies find that a boost of a factor of a few above electron scattering is reasonable in our temperature and ionization ranges, suggesting that more self consistent treatments of the opacity will yield robust wind solutions.     

We have assumed here the wind is optically thin in the UV and correspondingly the UV radiation field is time-independent. This assumption may be violated by both attenuation and re-radiation effects. We see that parts of the flow are optically thick, and therefore obscuration of the outer parts of the wind may be important. PSK00 had shown that accounting for UV attenuation in the approximation that the radiation is attenuated by radial columns of gas did not qualitatively change their solutions. However, optical depth effects of the UV continuum should not be too challenging to account for using our new methods. Also we expect that X-rays, attenuated by the gas, may re-radiate in the UV, and enhance the driving flux. Further, as some of our solutions are episodic, correctly accounting for light travel time effects may be important. 

Reprocessing effects may also be important. Assuming local thermodynamic equilibrium, gas at temperature $T$ will emit radiation in a Planck distribution. We may therefore account for the number of UV and X-ray photons emitted by the gas. The UV photons will change the radiation force by changing the local radiation flux. Likewise, the X-ray photons will alter the ionization state. As this work shows, the success of line driving for launching outflows is crucially related to the gas ionization so determining the ionization state and force multiplier are key. 

Monte Carlo simulations in the context of CVs have shown that ionization effects can severely inhibit line driving, suppressing or even preventing a wind (Sim10, \cite{Higginbottom14}). Their work solved for the radiation field, assuming that the hydrodynamics was fixed. Here, we find that, indeed the scattering can weaken a line driven wind, but the wind is still launched and accelerated due to its ability to adjust in a time-dependent fashion to a given radiation environment. A key lesson from their work is that suppression of the force multiplier is not due only to X-rays but also the FUV (Higginbottom et al. submitted to MNRAS). It highlights the non-linearity of these processes and the limitation of using the mean photon energy i.e ionization parameter approach. The shape of the SED, as also demonstrated in photoionization studies \cite{Dannen19} plays a crucial role in determining the force multiplier. Further, close to the threshold luminosities where line driving can operate as an acceleration mechanism, $\Gamma \sim 1/M_{\rm{max}}$, the mass outflow rate is strongly dependent on the maximum value of the force multiplier.  This work emphasises the need to correctly model the SED, and hence the radiation transfer, throughout the hydro simulation, while simultaneously computing the effects of this SED on the bulk gas properties such as force multiplier and heating and cooling.

Our study further highlights the need for a more self-consistent disc modeling. We have assumed a constant accretion rate of $1.3 M_{\odot}$/yr, and generate non-stationary winds with $\gtrsim 10\%$ this mass flux. This should have a non-negligible feedback on the disc, both in terms of changing the local accretion rate and altering the local radiation flux. Thus far, simulations have relied on semi-analytic \cite{Ganguly20}, \cite{Kirilov2023} or iterative methods \cite{Nomura2020} to account for this wind driven mass loss. Such large mass loss rates may have important implications for the accretion disk structure, possibly contributing to the mismatch between observed spectra and predictions of standard thin disk models \citep{SloneNetzer2012,LaorDavis2014,Nomura2020}.

Furthermore the contours of $E_r$ shown in Figure~\ref{fig:rad_field_summary} indicate that the optical depth in the radial direction near the disk is large when the outflow is present.  This will impact the self-irradiation of the disk and lead to a different reprocessing geometry than if the wind material was absent. This may help to explain the relatively large sizes inferred for optical/UV emission regions from continuum reverberation mapping.

In the future, we will build on these models by incorporating more realistic treatment of absorption, scattering and reprocessing. Using photoionization modeling, we can generate absorption opacity tables as a function of temperature and ionization parameter. The relative importance of radiative processes can then be accounted for in a self-consistent fashion, rather than the parameter survey type treatment we have used in this work.

\section*{Acknowledgments} 
Support for this work was provided by the National Aeronautics and Space Administration under TCAN grant 80NSSC21K0496. We thank Tim Kallman, Jim Stone, Yan-Fei Jiang and the entire DAWN TCAN collaboration for fruitful discussions. Further thanks to Nick Higginbottom, Nico Scepi, Christian Knigge, James Mattherws and Knox Long for sharing their recent Monte Carlo work. This work made use of Camber for cloud-based MPI provisioning.

\bibliographystyle{mnras}
\bibliography{progalab-shared}

\begin{thebibliography}{}
\makeatletter
\relax
\def\mn@urlcharsother{\let\do\@makeother \do\$\do\&\do\#\do\^\do\_\do\%\do\~}
\def\mn@doi{\begingroup\mn@urlcharsother \@ifnextchar [ {\mn@doi@}
  {\mn@doi@[]}}
\def\mn@doi@[#1]#2{\def\@tempa{#1}\ifx\@tempa\@empty \href
  {http://dx.doi.org/#2} {doi:#2}\else \href {http://dx.doi.org/#2} {#1}\fi
  \endgroup}
\def\mn@eprint#1#2{\mn@eprint@#1:#2::\@nil}
\def\mn@eprint@arXiv#1{\href {http://arxiv.org/abs/#1} {{\tt arXiv:#1}}}
\def\mn@eprint@dblp#1{\href {http://dblp.uni-trier.de/rec/bibtex/#1.xml}
  {dblp:#1}}
\def\mn@eprint@#1:#2:#3:#4\@nil{\def\@tempa {#1}\def\@tempb {#2}\def\@tempc
  {#3}\ifx \@tempc \@empty \let \@tempc \@tempb \let \@tempb \@tempa \fi \ifx
  \@tempb \@empty \def\@tempb {arXiv}\fi \@ifundefined
  {mn@eprint@\@tempb}{\@tempb:\@tempc}{\expandafter \expandafter \csname
  mn@eprint@\@tempb\endcsname \expandafter{\@tempc}}}

\bibitem[\protect\citeauthoryear{{Abbott}}{{Abbott}}{1982}]{Abbott1982}
{Abbott} D.~C.,  1982, \mn@doi [\apj] {10.1086/160166}, \href
  {https://ui.adsabs.harvard.edu/abs/1982ApJ...259..282A} {259, 282}

\bibitem[\protect\citeauthoryear{{Arav} \& {Li}}{{Arav} \&
  {Li}}{1994}]{Arav1994a}
{Arav} N.,  {Li} Z.-Y.,  1994, \mn@doi [\apj] {10.1086/174177}, \href
  {https://ui.adsabs.harvard.edu/abs/1994ApJ...427..700A} {427, 700}

\bibitem[\protect\citeauthoryear{{Arav}, {Li}  \& {Begelman}}{{Arav}
  et~al.}{1994}]{Arav1994b}
{Arav} N.,  {Li} Z.-Y.,   {Begelman} M.~C.,  1994, \mn@doi [\apj]
  {10.1086/174549}, \href
  {https://ui.adsabs.harvard.edu/abs/1994ApJ...432...62A} {432, 62}

\bibitem[\protect\citeauthoryear{{Blondin}}{{Blondin}}{1994}]{Blondin1994}
{Blondin} J.~M.,  1994, \mn@doi [\apj] {10.1086/174853}, \href
  {https://ui.adsabs.harvard.edu/abs/1994ApJ...435..756B} {435, 756}

\bibitem[\protect\citeauthoryear{{Blondin}, {Kallman}, {Fryxell}  \&
  {Taam}}{{Blondin} et~al.}{1990}]{Blondin1990}
{Blondin} J.~M.,  {Kallman} T.~R.,  {Fryxell} B.~A.,   {Taam} R.~E.,  1990,
  \mn@doi [\apj] {10.1086/168865}, \href
  {https://ui.adsabs.harvard.edu/abs/1990ApJ...356..591B} {356, 591}

\bibitem[\protect\citeauthoryear{{Castor}, {Abbott}  \& {Klein}}{{Castor}
  et~al.}{1975}]{CAK1975}
{Castor} J.~I.,  {Abbott} D.~C.,   {Klein} R.~I.,  1975, \mn@doi [\apj]
  {10.1086/153315}, \href
  {https://ui.adsabs.harvard.edu/abs/1975ApJ...195..157C} {195, 157}

\bibitem[\protect\citeauthoryear{{Castro Segura} et~al.,}{{Castro Segura}
  et~al.}{2022}]{CastroSegura2022}
{Castro Segura} N.,  et~al., 2022, \mn@doi [\nat] {10.1038/s41586-021-04324-2},
  \href {https://ui.adsabs.harvard.edu/abs/2022Natur.603...52C} {603, 52}

\bibitem[\protect\citeauthoryear{{Dannen}, {Proga}, {Kallman}  \&
  {Waters}}{{Dannen} et~al.}{2019}]{Dannen19}
{Dannen} R.~C.,  {Proga} D.,  {Kallman} T.~R.,   {Waters} T.,  2019, \mn@doi
  [\apj] {10.3847/1538-4357/ab340b}, \href
  {https://ui.adsabs.harvard.edu/abs/2019ApJ...882...99D} {882, 99}

\bibitem[\protect\citeauthoryear{{Dannen}, {Proga}  \& {Waters}}{{Dannen}
  et~al.}{2023}]{Dannen2023}
{Dannen} R.,  {Proga} D.,   {Waters} T.,  2023, \mn@doi [arXiv e-prints]
  {10.48550/arXiv.2306.04063}, \href
  {https://ui.adsabs.harvard.edu/abs/2023arXiv230604063D} {p. arXiv:2306.04063}

\bibitem[\protect\citeauthoryear{{Dyda} \& {Proga}}{{Dyda} \&
  {Proga}}{2018a}]{DP2018a}
{Dyda} S.,  {Proga} D.,  2018a, \mn@doi [\mnras] {10.1093/mnras/sty030}, \href
  {https://ui.adsabs.harvard.edu/abs/2018MNRAS.475.3786D} {475, 3786}

\bibitem[\protect\citeauthoryear{{Dyda} \& {Proga}}{{Dyda} \&
  {Proga}}{2018b}]{DP2018b}
{Dyda} S.,  {Proga} D.,  2018b, \mn@doi [\mnras] {10.1093/mnras/sty1257}, \href
  {https://ui.adsabs.harvard.edu/abs/2018MNRAS.478.5006D} {478, 5006}

\bibitem[\protect\citeauthoryear{{Ganguly} \& {Proga}}{{Ganguly} \&
  {Proga}}{2020}]{Ganguly20}
{Ganguly} S.,  {Proga} D.,  2020, \mn@doi [\apj] {10.3847/1538-4357/ab6aa0},
  \href {https://ui.adsabs.harvard.edu/abs/2020ApJ...890...54G} {890, 54}

\bibitem[\protect\citeauthoryear{{Gayley}}{{Gayley}}{1995}]{Gayley1995}
{Gayley} K.~G.,  1995, \mn@doi [\apj] {10.1086/176492}, \href
  {https://ui.adsabs.harvard.edu/abs/1995ApJ...454..410G} {454, 410}

\bibitem[\protect\citeauthoryear{{Higginbottom}, {Proga}, {Knigge}, {Long},
  {Matthews}  \& {Sim}}{{Higginbottom} et~al.}{2014}]{Higginbottom14}
{Higginbottom} N.,  {Proga} D.,  {Knigge} C.,  {Long} K.~S.,  {Matthews} J.~H.,
    {Sim} S.~A.,  2014, \mn@doi [\apj] {10.1088/0004-637X/789/1/19}, \href
  {https://ui.adsabs.harvard.edu/abs/2014ApJ...789...19H} {789, 19}

\bibitem[\protect\citeauthoryear{{Higginbottom}, {Proga}, {Knigge}  \&
  {Long}}{{Higginbottom} et~al.}{2017}]{Higginbottom17}
{Higginbottom} N.,  {Proga} D.,  {Knigge} C.,   {Long} K.~S.,  2017, \mn@doi
  [\apj] {10.3847/1538-4357/836/1/42}, \href
  {https://ui.adsabs.harvard.edu/abs/2017ApJ...836...42H} {836, 42}

\bibitem[\protect\citeauthoryear{{Higginbottom}, {Knigge}, {Sim}, {Long},
  {Matthews}, {Hewitt}, {Parkinson}  \& {Mangham}}{{Higginbottom}
  et~al.}{2020}]{Higginbottom2020}
{Higginbottom} N.,  {Knigge} C.,  {Sim} S.~A.,  {Long} K.~S.,  {Matthews}
  J.~H.,  {Hewitt} H.~A.,  {Parkinson} E.~J.,   {Mangham} S.~W.,  2020, \mn@doi
  [\mnras] {10.1093/mnras/staa209}, \href
  {https://ui.adsabs.harvard.edu/abs/2020MNRAS.492.5271H} {492, 5271}

\bibitem[\protect\citeauthoryear{{Jiang}}{{Jiang}}{2021}]{Jiang2021}
{Jiang} Y.-F.,  2021, \mn@doi [\apjs] {10.3847/1538-4365/abe303}, \href
  {https://ui.adsabs.harvard.edu/abs/2021ApJS..253...49J} {253, 49}

\bibitem[\protect\citeauthoryear{{Jiang}}{{Jiang}}{2022}]{Jiang2022}
{Jiang} Y.-F.,  2022, \mn@doi [\apjs] {10.3847/1538-4365/ac9231}, \href
  {https://ui.adsabs.harvard.edu/abs/2022ApJS..263....4J} {263, 4}

\bibitem[\protect\citeauthoryear{{Kirilov}, {Dyda}  \& {Reynolds}}{{Kirilov}
  et~al.}{2023}]{Kirilov2023}
{Kirilov} A.,  {Dyda} S.,   {Reynolds} C.~S.,  2023, \mn@doi [\mnras]
  {10.1093/mnras/stad083}, \href
  {https://ui.adsabs.harvard.edu/abs/2023MNRAS.520...44K} {520, 44}

\bibitem[\protect\citeauthoryear{{Laor} \& {Davis}}{{Laor} \&
  {Davis}}{2014}]{LaorDavis2014}
{Laor} A.,  {Davis} S.~W.,  2014, \mn@doi [\mnras] {10.1093/mnras/stt2408},
  \href {https://ui.adsabs.harvard.edu/abs/2014MNRAS.438.3024L} {438, 3024}

\bibitem[\protect\citeauthoryear{{Lucy} \& {Solomon}}{{Lucy} \&
  {Solomon}}{1970}]{Lucy1970}
{Lucy} L.~B.,  {Solomon} P.~M.,  1970, \mn@doi [\apj] {10.1086/150365}, \href
  {https://ui.adsabs.harvard.edu/abs/1970ApJ...159..879L} {159, 879}

\bibitem[\protect\citeauthoryear{{Mata S{\'a}nchez} et~al.,}{{Mata S{\'a}nchez}
  et~al.}{2018}]{Mata-Sanchez2018}
{Mata S{\'a}nchez} D.,  et~al., 2018, \mn@doi [\mnras] {10.1093/mnras/sty2402},
  \href {https://ui.adsabs.harvard.edu/abs/2018MNRAS.481.2646M} {481, 2646}

\bibitem[\protect\citeauthoryear{{Matthews} et~al.,}{{Matthews}
  et~al.}{2023}]{Matthews2023}
{Matthews} J.~H.,  et~al., 2023, \mn@doi [\mnras] {10.1093/mnras/stad2895},
  \href {https://ui.adsabs.harvard.edu/abs/2023MNRAS.tmp.2847M} {}

\bibitem[\protect\citeauthoryear{{Miller} et~al.,}{{Miller}
  et~al.}{2016}]{Miller2016}
{Miller} J.~M.,  et~al., 2016, \mn@doi [\apjl] {10.3847/2041-8205/821/1/L9},
  \href {https://ui.adsabs.harvard.edu/abs/2016ApJ...821L...9M} {821, L9}

\bibitem[\protect\citeauthoryear{{Mu{\~n}oz-Darias} et~al.,}{{Mu{\~n}oz-Darias}
  et~al.}{2019}]{Munoz-Darias2019}
{Mu{\~n}oz-Darias} T.,  et~al., 2019, \mn@doi [\apjl]
  {10.3847/2041-8213/ab2768}, \href
  {https://ui.adsabs.harvard.edu/abs/2019ApJ...879L...4M} {879, L4}

\bibitem[\protect\citeauthoryear{{Murray}, {Chiang}, {Grossman}  \&
  {Voit}}{{Murray} et~al.}{1995}]{Murray1995}
{Murray} N.,  {Chiang} J.,  {Grossman} S.~A.,   {Voit} G.~M.,  1995, \mn@doi
  [\apj] {10.1086/176238}, \href
  {https://ui.adsabs.harvard.edu/abs/1995ApJ...451..498M} {451, 498}

\bibitem[\protect\citeauthoryear{{Nomura} \& {Ohsuga}}{{Nomura} \&
  {Ohsuga}}{2017}]{Nomura2017}
{Nomura} M.,  {Ohsuga} K.,  2017, \mn@doi [\mnras] {10.1093/mnras/stw2877},
  \href {https://ui.adsabs.harvard.edu/abs/2017MNRAS.465.2873N} {465, 2873}

\bibitem[\protect\citeauthoryear{{Nomura}, {Ohsuga}, {Takahashi}, {Wada}  \&
  {Yoshida}}{{Nomura} et~al.}{2016}]{Nomura2016}
{Nomura} M.,  {Ohsuga} K.,  {Takahashi} H.~R.,  {Wada} K.,   {Yoshida} T.,
  2016, \mn@doi [\pasj] {10.1093/pasj/psv124}, \href
  {https://ui.adsabs.harvard.edu/abs/2016PASJ...68...16N} {68, 16}

\bibitem[\protect\citeauthoryear{{Nomura}, {Ohsuga}  \& {Done}}{{Nomura}
  et~al.}{2020}]{Nomura2020}
{Nomura} M.,  {Ohsuga} K.,   {Done} C.,  2020, \mn@doi [\mnras]
  {10.1093/mnras/staa948}, \href
  {https://ui.adsabs.harvard.edu/abs/2020MNRAS.494.3616N} {494, 3616}

\bibitem[\protect\citeauthoryear{{Owocki}, {Castor}  \& {Rybicki}}{{Owocki}
  et~al.}{1988}]{Owocki1988}
{Owocki} S.~P.,  {Castor} J.~I.,   {Rybicki} G.~B.,  1988, \mn@doi [\apj]
  {10.1086/166977}, \href
  {https://ui.adsabs.harvard.edu/abs/1988ApJ...335..914O} {335, 914}

\bibitem[\protect\citeauthoryear{{Pereyra}, {Kallman}  \& {Blondin}}{{Pereyra}
  et~al.}{1997}]{Pereyra1997}
{Pereyra} N.~A.,  {Kallman} T.~R.,   {Blondin} J.~M.,  1997, \mn@doi [\apj]
  {10.1086/303671}, \href
  {https://ui.adsabs.harvard.edu/abs/1997ApJ...477..368P} {477, 368}

\bibitem[\protect\citeauthoryear{{Proga}}{{Proga}}{2003}]{Proga2003}
{Proga} D.,  2003, \mn@doi [\apj] {10.1086/345897}, \href
  {https://ui.adsabs.harvard.edu/abs/2003ApJ...585..406P} {585, 406}

\bibitem[\protect\citeauthoryear{{Proga} \& {Kallman}}{{Proga} \&
  {Kallman}}{2002}]{Proga2002}
{Proga} D.,  {Kallman} T.~R.,  2002, \mn@doi [\apj] {10.1086/324534}, \href
  {https://ui.adsabs.harvard.edu/abs/2002ApJ...565..455P} {565, 455}

\bibitem[\protect\citeauthoryear{{Proga} \& {Kallman}}{{Proga} \&
  {Kallman}}{2004}]{PK04}
{Proga} D.,  {Kallman} T.~R.,  2004, \mn@doi [\apj] {10.1086/425117}, \href
  {https://ui.adsabs.harvard.edu/abs/2004ApJ...616..688P} {616, 688}

\bibitem[\protect\citeauthoryear{{Proga}, {Stone}  \& {Drew}}{{Proga}
  et~al.}{1998}]{PSD98}
{Proga} D.,  {Stone} J.~M.,   {Drew} J.~E.,  1998, \mn@doi [\mnras]
  {10.1046/j.1365-8711.1998.01337.x}, \href
  {https://ui.adsabs.harvard.edu/abs/1998MNRAS.295..595P} {295, 595}

\bibitem[\protect\citeauthoryear{{Proga}, {Stone}  \& {Drew}}{{Proga}
  et~al.}{1999}]{PSD99}
{Proga} D.,  {Stone} J.~M.,   {Drew} J.~E.,  1999, \mn@doi [\mnras]
  {10.1046/j.1365-8711.1999.02935.x}, \href
  {https://ui.adsabs.harvard.edu/abs/1999MNRAS.310..476P} {310, 476}

\bibitem[\protect\citeauthoryear{{Proga}, {Stone}  \& {Kallman}}{{Proga}
  et~al.}{2000}]{PSK2000}
{Proga} D.,  {Stone} J.~M.,   {Kallman} T.~R.,  2000, \mn@doi [\apj]
  {10.1086/317154}, \href
  {https://ui.adsabs.harvard.edu/abs/2000ApJ...543..686P} {543, 686}

\bibitem[\protect\citeauthoryear{{Schurch}, {Done}  \& {Proga}}{{Schurch}
  et~al.}{2009}]{Schurch2009}
{Schurch} N.~J.,  {Done} C.,   {Proga} D.,  2009, \mn@doi [\apj]
  {10.1088/0004-637X/694/1/1}, \href
  {https://ui.adsabs.harvard.edu/abs/2009ApJ...694....1S} {694, 1}

\bibitem[\protect\citeauthoryear{{Shakura} \& {Sunyaev}}{{Shakura} \&
  {Sunyaev}}{1973}]{SS73}
{Shakura} N.~I.,  {Sunyaev} R.~A.,  1973, \aap, \href
  {https://ui.adsabs.harvard.edu/abs/1973A&A....24..337S} {500, 33}

\bibitem[\protect\citeauthoryear{{Sim}, {Drew}  \& {Long}}{{Sim}
  et~al.}{2005}]{Sim2005}
{Sim} S.~A.,  {Drew} J.~E.,   {Long} K.~S.,  2005, \mn@doi [\mnras]
  {10.1111/j.1365-2966.2005.09472.x}, \href
  {https://ui.adsabs.harvard.edu/abs/2005MNRAS.363..615S} {363, 615}

\bibitem[\protect\citeauthoryear{{Sim}, {Long}, {Miller}  \& {Turner}}{{Sim}
  et~al.}{2008}]{Sim2008}
{Sim} S.~A.,  {Long} K.~S.,  {Miller} L.,   {Turner} T.~J.,  2008, \mn@doi
  [\mnras] {10.1111/j.1365-2966.2008.13466.x}, \href
  {https://ui.adsabs.harvard.edu/abs/2008MNRAS.388..611S} {388, 611}

\bibitem[\protect\citeauthoryear{{Sim}, {Proga}, {Miller}, {Long}  \&
  {Turner}}{{Sim} et~al.}{2010}]{Sim2010}
{Sim} S.~A.,  {Proga} D.,  {Miller} L.,  {Long} K.~S.,   {Turner} T.~J.,  2010,
  \mn@doi [\mnras] {10.1111/j.1365-2966.2010.17215.x}, \href
  {https://ui.adsabs.harvard.edu/abs/2010MNRAS.408.1396S} {408, 1396}

\bibitem[\protect\citeauthoryear{{Slone} \& {Netzer}}{{Slone} \&
  {Netzer}}{2012}]{SloneNetzer2012}
{Slone} O.,  {Netzer} H.,  2012, \mn@doi [\mnras]
  {10.1111/j.1365-2966.2012.21699.x}, \href
  {https://ui.adsabs.harvard.edu/abs/2012MNRAS.426..656S} {426, 656}

\bibitem[\protect\citeauthoryear{{Stevens} \& {Kallman}}{{Stevens} \&
  {Kallman}}{1990}]{Stevens1990}
{Stevens} I.~R.,  {Kallman} T.~R.,  1990, \mn@doi [\apj] {10.1086/169486},
  \href {https://ui.adsabs.harvard.edu/abs/1990ApJ...365..321S} {365, 321}

\bibitem[\protect\citeauthoryear{{Tomaru}, {Done}, {Ohsuga}, {Odaka}  \&
  {Takahashi}}{{Tomaru} et~al.}{2020}]{Tomaru2020}
{Tomaru} R.,  {Done} C.,  {Ohsuga} K.,  {Odaka} H.,   {Takahashi} T.,  2020,
  \mn@doi [\mnras] {10.1093/mnras/staa961}, \href
  {https://ui.adsabs.harvard.edu/abs/2020MNRAS.494.3413T} {494, 3413}

\bibitem[\protect\citeauthoryear{{White}, {Mullen}, {Jiang}, {Davis}, {Stone},
  {Morozova}  \& {Zhang}}{{White} et~al.}{2023}]{Whiteetal2023}
{White} C.~J.,  {Mullen} P.~D.,  {Jiang} Y.-F.,  {Davis} S.~W.,  {Stone} J.~M.,
   {Morozova} V.,   {Zhang} L.,  2023, \mn@doi [\apj]
  {10.3847/1538-4357/acc8cf}, \href
  {https://ui.adsabs.harvard.edu/abs/2023ApJ...949..103W} {949, 103}

\bibitem[\protect\citeauthoryear{{de Kool} \& {Begelman}}{{de Kool} \&
  {Begelman}}{1995}]{deKool1995}
{de Kool} M.,  {Begelman} M.~C.,  1995, \mn@doi [\apj] {10.1086/176594}, \href
  {https://ui.adsabs.harvard.edu/abs/1995ApJ...455..448D} {455, 448}

\makeatother
\end{thebibliography}

\bsp	
\label{lastpage}
\end{document}